\documentclass{article}
\usepackage{spconf,amsmath,graphicx}
\usepackage{amsfonts}
\usepackage{multirow}
\usepackage{makecell}
\usepackage{svg}
\usepackage{dirtree}
\PassOptionsToPackage{hyphens}{url}\usepackage{hyperref}
\hypersetup{colorlinks=true}
\usepackage{cleveref}
\usepackage{xcolor}

\title{Sharing Low Rank Conformer Weights for Tiny Always-On Ambient Speech Recognition Models}
\name{Steven M. Hernandez$^{1}$, Ding Zhao$^{2}$, Shaojin Ding$^{2}$, Antoine Bruguier$^{2}$,}
\nameplus{Rohit Prabhavalkar$^{2}$, Tara N. Sainath$^{2}$, Yanzhang He$^{2}$, Ian McGraw$^{2}$\thanks{This work was done while Steven M. Hernandez was an intern at Google.}}
\address{$^{1}$Virginia Commonwealth University, $^{2}$Google}
\begin{document}
\ninept
\maketitle
\begin{abstract}
Continued improvements in machine learning techniques offer exciting new opportunities through the use of larger models and larger training datasets. However, there is a growing need to offer these new capabilities on-board low-powered devices such as smartphones, wearables and other embedded environments where only low memory is available. Towards this, we consider methods to reduce the model size of Conformer-based speech recognition models which typically require models with greater than $100$M parameters down to just $5$M parameters while minimizing impact on model quality. Such a model allows us to achieve always-on ambient speech recognition on edge devices with low-memory neural processors. We propose model weight reuse at different levels within our model architecture: (i)~repeating full conformer block layers, (ii)~sharing specific conformer modules across layers, (iii)~sharing sub-components per conformer module, and (iv)~sharing decomposed sub-component weights after low-rank decomposition. By sharing weights at different levels of our model, we can retain the full model in-memory 
while  increasing the number of virtual transformations applied to the input. Through a series of ablation studies and evaluations, we find that with weight sharing and a low-rank architecture, we can achieve a WER of 2.84 and 2.94 for Librispeech dev-clean and test-clean respectively with a $5$M parameter model.

\end{abstract}

\section{Introduction}
\label{sec:introduction}
\vspace{-3pt}

Automatic speech recognition (ASR) is an essential component in a growing number of spoken language interfaces on mobile devices. Recently, long running applications such as transcribed recording, live captioning, and generalized keyword spotting are emerging, and are even more challenging on edge devices due to the limited resources. Always-on ambient speech recognition, the most ambitious use scenario, leverages advances in both deep learning and embedded neural processing hardware to enable always-running ASR on low power edge devices.

To achieve efficient always-on recognition with edge devices, instead of running a standard ASR model, alternative approaches are usually considered, such as recognizing a single keyword~\cite{alvarez2019end} or just a small set of intents~\cite{ray2021listen}. 
However, this is not possible for keywordless interactions and also requires newly trained models each time new intents are made available. As such, in this work, we aim to focus on the extendability achieved by a generalized speech recognition model.
However, recent advances in machine learning come at the expense of ever increasing model sizes (i.e., $100$M parameters and higher). These models are not tractable for always-running on neural accelerators such as edge TPUs~\cite{antonini2019resource}, which are limited to fewer than $6$M parameters due to hardware memory constraints. In fact, to achieve inference using such large model sizes, the model must be split into smaller chunks which are then continuously transferred from memory to TPU, leading to poor energy usage and poor latency for ambient speech recognition tasks.

In this paper, we look for methods to reduce the size of Conformer-based~\cite{conformer} speech recognition models to achieve always-on ambient speech recognition, which can efficiently leverage specialized hardware such as edge TPUs. To do this, we propose model weight reuse at different levels within our Conformer architecture such as: (i) repeating full conformer layers, (ii) sharing specific modules across conformer layers, (iii) sharing specific sub-components within each conformer module, and (iv) sharing low-rank sub-weights after low-rank decomposition. 
Unlike other model compression techniques like low-bit quantization~\cite{hubara2016binarized} and sparsity~\cite{han2015learning} which assume the use of specialized hardware features which we will discuss in \Cref{sec:related_works}, both sharing and low rank architectures can be achieved with existing neural accelerators.
By sharing weights across layers, we can increase the number of virtual transformations applied to our input data without increasing the physical size of the model weights in memory. 
Increasing the number of virtual transformations in our model allows for 
more complex
transformations on our model input which emulates the transforms typically found by increasing the number of layers in a model.

\section{Related Works}
\label{sec:related_works}
\vspace{-3pt}

Performing machine learning model inference on-board low power edge devices has recently achieved greater attention for tasks such as device-free wireless sensing~\cite{hernandez2022wifi}, computer vision~\cite{xu2022ultra}, and numerous other tasks~\cite{banbury2020benchmarking}. 
At the core of edge model inference is achieving model compression for use on low power and low resourced devices.

Model compression has commonly been achieved through a number of methods such as 
sparsity pruning~\cite{han2015learning,wu2021dynamic,ding2021audio}, low-bit quantization~\cite{zhou2018adaptive, novac2021quantization, ding22c_interspeech}, %
knowledge distillation~\cite{DBLP:conf/interspeech/CeruttiPBF19, yang2020model}, and low-rank matrix factorization~\cite{DBLP:journals/corr/abs-2207-00112,yu2017compressing}.
These techniques can typically be applied regardless of the model architecture which allows them to be generalized to different tasks. However, some methods assume access to specific hardware features that may not be available on edge devices.
Model sparsity techniques offers the ability to prune weights until an exact model size is achieved. However, without structured sparsity~\cite{wen2016learning}, the resulting model requires irregular memory access and without hardware support, memory usage and computation become inefficient.
Quantization is typically applied to reduce model weights from $32$-bit floating point values down to $8$-bit integer values, and is also applied to lower quantization levels (i.e., $1$-bit, $2$-bit, or $4$-bit~\cite{hubara2016binarized, ding22c_interspeech}) and even mixed-precision quantization~\cite{schaefer2022edge}.
However, computations on low-bit quantization level models are not available on typical real-world hardware.
On the other hand, techniques like knowledge distillation and low-rank decomposition are computed off-device and thus performing inference on these compressed models 
is identical to non-compressed models.

\vspace{-6pt}
\section{Methods}
\label{sec:methods}
\vspace{-3pt}

\subsection{Conformer Model}

\vspace{-3pt}
For our ambient ASR task, we leverage the conformer model architecture~\cite{conformer}, an extension to the transformer model architecture~\cite{vaswani2017attention}. For the intents of this work, we will focus on reducing the size of the conformer encoder since we find that it takes up greater than $90\%$ of the overall model size. The size of the encoder is primarily a result of the $N$ conformer blocks, thus we will also focus on ways to reduce the size of the encoder by both reducing the size of the individual conformer blocks as well as reducing the need for large values of $N$.

We define the $i$-th conformer block $\mathbb{C}_{(i)}$ in our model as $\mathbb{C}_{(i)}\big(F_\text{start}^{(i)}, A^{(i)}, C^{(i)}, F_\text{end}^{(i)}\big)$ where $F_\text{start}$, $A$, $C$, $F_\text{end}$ are the parameters for the feed forward start, attention, convolution and, feed forward end modules respectively within the $i$-th conformer block $\mathbb{C}_{(i)}$. 
Fig.~\ref{fig:conformer_block_structure} illustrates the largest size sub-components for each of the modules. It is important to recognize these largest sub-components when reducing the size of our model because these are the weight matrices which we should focus on compressing. Notice, that while the architecture of our conformer contains many unique features, %
the size of the conformer blocks is primarily the result of several linear layers.
Thus, if we can apply a compression technique to simple linear layers, then they 
can be
similarly applied throughout the entire conformer model.

\vspace{-3pt}
\subsection{Repeat Full Layers}
\vspace{-2pt}

Suppose we have a conformer model with $N$ conformer blocks $\mathbb{C}_{(i)}$ where $i \in \{1, \dots, N\}$, then we can repeat each $i$-th layer $R[i]$ times as suggested in~\cite{dabre2019recurrent} by sharing the $i$-th layer's parameters. Our conformer transformation will then be described as a series of $n$-fold iterative functions defined as:

\begin{equation}
    f^{\circ n} = \underbrace{f \circ f \circ \dots \circ f}_{n},
\end{equation}
where $f$ is some transformation function, and $n$ is the number of times that the function is composed over itself. As such, our conformer transformation could be described as:
\vspace{-5pt}
\begin{equation}
    \underbrace{\mathbb{C}_{(N)} \circ \dots \circ \mathbb{C}_{(N)}}_{R[N]} \circ  \dots \circ
    \underbrace{\mathbb{C}_{(2)} \circ \dots \circ \mathbb{C}_{(2)}}_{R[2]} \circ
    \underbrace{\mathbb{C}_{(1)} \circ \dots \circ \mathbb{C}_{(1)}}_{R[1]},
\end{equation}
\vspace{-5pt}
or 
\begin{equation}
    \mathbb{C}_{(N)}^{\circ R[N]} \circ 
    \dots \circ 
    \mathbb{C}_{(2)}^{\circ R[2]} \circ 
    \mathbb{C}_{(1)}^{\circ R[1]}.
\end{equation}
By repeating layers, we retain a set number of physical conformer layers ($N$) while increasing the number of virtual conformer layers or the number of transformations ($R \times N$). By performing $n$-fold iterative conformer transformations, we expect that we can transform our input with more complex transformations without directly increasing our cost from a model size perspective. 

\vspace{-3pt}
\subsection{Sub-component Customization}
\vspace{-2pt}
While we expect that the higher complexity transformations offered by repeating layers will be of benefit to our model architecture, there is still a clear intuition that a model with $R \times N$ virtual layers will likely not be able to perform better than a model with $R \times N$ physical layers due to the increased number of distinct parameters available in the model. Thus, we may wish to allow for layer repeating but with some slight customization per layer. 
By allowing customization, we can retain our ability to reduce model size through sharing, but we also allow conformer blocks to perform unique transformations instead of strictly iterative function composition.

Our first effort towards this is to only share certain modules within our conformer blocks. 
Sharing specific modules such as feed forward or attention modules was initially reviewed in~\cite{lan2019albert} towards a reduced size BERT model. We define sharing indices $\mathcal{I}_{FS}$, $\mathcal{I}_{A}$, $\mathcal{I}_{C}$, $\mathcal{I}_{FE}$ which correspond to our feed forward start, attention, convolution, and feed forward end modules respectively\footnote{Other modules are ignored due to the large relative size of these modules.}. Each of the described sharing indices $\mathcal{I}_{x}$, are subject to the following constraints: 
$|\mathcal{I}_{x}| = N$, 
$min(\mathcal{I}_{x}) = 1$, 
$max(\mathcal{I}_{x}) \leq N$. 
With these sharing indices, we can now define our $i$-th conformer block as: 
\vspace{-3pt}
\begin{equation}
\mathbb{C}_{(i)}\bigg(
F_{start}^{\big(\mathcal{I}_{FS}^{(i)}\big)}, 
A^{\big(\mathcal{I}_{A}^{(i)}\big)}, 
C^{\big(\mathcal{I}_{C}^{(i)}\big)}, 
F_{end}^{\big(\mathcal{I}_{FE}^{(i)}\big)}
\bigg).
\vspace{-5pt}
\end{equation}

\begin{figure}
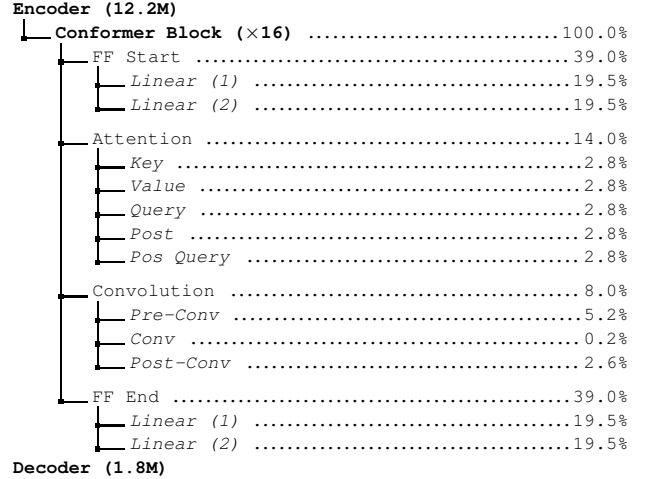

    \centering

\begin{minipage}{\linewidth}
\setlength{\DTbaselineskip}{9pt}
\scriptsize
\dirtree{%
.1 \textbf{Encoder (12.2M)}.
.2 \textbf{Conformer Block ($\times$16)} \dotfill 100.0\%.
.3 \text{FF Start} \dotfill 39.0\%.
.4 \textit{Linear (1)} \dotfill 19.5\%.
.4 \textit{Linear (2)} \dotfill 19.5\%\vspace{1mm}.
.3 \text{Attention} \dotfill 14.0\%.
.4 \textit{Key} \dotfill 2.8\%.
.4 \textit{Value} \dotfill 2.8\%.
.4 \textit{Query} \dotfill 2.8\%.
.4 \textit{Post} \dotfill 2.8\%.
.4 \textit{Pos Query} \dotfill 2.8\%\vspace{1mm}.
.3 \text{Convolution} \dotfill 8.0\%.
.4 \textit{Pre-Conv} \dotfill 5.2\%.
.4 \textit{Conv} \dotfill 0.2\%.
.4 \textit{Post-Conv} \dotfill 2.6\%\vspace{1mm}.
.3 \text{FF End} \dotfill 39.0\%.
.4 \textit{Linear (1)} \dotfill 19.5\%.
.4 \textit{Linear (2)} \dotfill 19.5\%.
}
\dirtree{%
.1 \textbf{Decoder (1.8M)}.
}
\end{minipage}
    
    \caption{High level composition of our base $14$M parameter Conformer model (B0). We focus on reducing the size of the $16$ conformer blocks within the encoder portion through sharing 
    (i) full conformer blocks, (ii) modules, and (iii) sub-components.
    }
    \label{fig:conformer_block_structure}
\vspace{-15pt}
\end{figure}

Digging deeper into the structure of our conformer modules, even smaller sub-components can be found. In Fig.~\ref{fig:conformer_block_structure} we see that the largest sub-components are primarily linear layers. Beyond these sub-components, there are also some much smaller modules which have miniscule effect on decreasing the model size, yet may allow for better customization of our shared conformer layers. As such, we suggested that these smaller components can be excluded from our model sharing system, thus allowing improved model performance. Furthermore, we expect that certain sub-components may hold great importance in providing improved model performance and thus should not be shared even though they do
contribute to increasing model size.

\subsection{Low-Rank Factorization}

As illustrated in Fig.~\ref{fig:conformer_block_structure}, the largest sub-components within our conformer are normal linear layers composed of a weight matrix $M$ and bias $b$. The complexity of the conformer comes from the specific architecture of the model rather than the complexity of the specific sub-components. Since the weight matrix is much larger in size than the bias, we will focus on reducing the size of $M$.
Supposing that $M \in \mathbb{R}^{m \times n}$, we can apply low-rank decomposition~\cite{DBLP:journals/corr/abs-2207-00112} to reduce $M$ into three distinct sub-matrices: $U \in \mathbb{R}^{m \times k}$, $V \in \mathbb{R}^{n \times k}$, and $\Sigma \in \mathbb{R}^{k \times k}$ a diagonal matrix which can be found through:
\vspace{-3pt}
\begin{equation}
\min_{U,\Sigma,V} ||M - U \Sigma V^T||,  %
\vspace{-3pt}
\end{equation}
using singular value decomposition (SVD). 
Given $k \ll \text{min}(m, n)$, 
the number of parameters for any $M$ in our model can be reduced.

\begin{table}[]
    \centering
    \begin{tabular}{|c|c|c|c|c|c|}
    \cline{2-6}
        \multicolumn{1}{c|}{} & \multicolumn{2}{c|}{\thead{\# Conformer Layers}} & \multicolumn{2}{c|}{\thead{WER}} & \multirow{2}{*}{\thead{Model\\Size}} \\
    \cline{2-5}
        \multicolumn{1}{c|}{} & \thead{Physical} & \thead{Virtual} & \thead{dev} & \thead{test} &  \\
\cline{2-6}\noalign{\vspace{2.5pt}}
    \hline
        SL0 & 1 & 1 & 10.69 & 10.53 & 2.55M \\
    \hline
        SL1 & 1 & 2 & 9.15 & 9.18 & 2.55M \\
    \hline
        SL2 & 1 & 3 & 8.39 & 8.71 & 2.55M \\
    \hline 
    \hline
        SL3 & 4 & 4 & 4.13 & 4.30 & 4.84M \\
    \hline
        SL4 & 4 & 8 & 3.50 & 3.76 & 4.84M \\
    \hline
        SL5 & 4 & 12 & 3.20 & 3.45 & 4.84M \\
    \hline
        SL6 & 4 & 16 & 3.31 & 3.67 & 4.84M \\
    \hline 
    \end{tabular}
    \caption{Sharing full layers by passing layer output into itself.}
    \label{tab:share_conformer_blocks}
\vspace{-10pt}
\end{table}

While reducing $k$ allows us to reduce the number of parameters for a given matrix, it also greatly increases the reconstruction error. To account for this, typically model fine-tuning is performed after decomposition to account for loss in model performance. In our case, we instead begin with a low-rank structure when training from scratch which allows us to forgo the need for training followed by fine tuning and also allows us to ignore the use of singular value decomposition. As such, for simplicity, $\Sigma$ can be combined with $U$ or $V$ in our low-rank reconstruction structure as suggested in~\cite{xue2013restructuring}. Thus, we can restate our reconstruction structure as $M \sim U V^T$.

\vspace{-3pt}
\section{Experiment Design}

\vspace{-3pt}
\subsection{Dataset}

We evaluate on the LibriSpeech datasets~\cite{panayotov2015librispeech} which consists of $960$ hours of training data (i.e., train-clean and train-other)
and we evaluate on dev-clean and test-clean.
The spoken-word input data is structured as $80$ log Mel-filterbank energy features with a window size of $25$ms and a $10$ms stride. The output is modelled using a word-piece model (WPM)
embedding with a dimensionality of $1,024$. 

\vspace{-3pt}
\subsection{Model Architecture}

We begin our evaluations with a conformer architecture (B0) with $14$M parameters and $16$ conformer block layers consisting of $0.7$M parameters each. 
This baseline architecture achieves a word-error rate (WER) of $2.18$ on dev-clean and $2.53$ on test-clean, however, our goal is to reduce this model down to approximately $5$M parameters (a reduction of approximately $-65\%$), thus this model is not applicable for always-on ambient ASR using low-power edge TPU devices.
Notice, we begin with a $14$M parameter model rather than a larger $100$M$+$ parameter model since it is a common size for ``small'' models in the literature~\cite{conformer, han2020contextnet}.
We also design a handcrafted $5$M parameter version of this model (B1) as a baseline with $8$ conformer block layers with a size of $0.5$M parameters each. This $5$M baseline model achieves a WER of $3.53$ on dev-clean
and $3.72$ on test-clean.
\vspace{-5pt}
\section{Results}
\label{sec:results}
\vspace{-3pt}

\subsection{Repeat Full Layers}
We begin our evaluations by reviewing the results of sharing full conformer layers. When sharing conformer layers, we repeat the conformer transformations in order over multiple repetitions. Suppose we have $N$ conformer blocks repeated $R$ times, we define the number of physical conformer layers as $N$ and the number of virtual conformer layers as $N \times R$. By increasing $R$, we can achieve an increase in the number of transformations applied to our input data with the expectation that increasing the number of transformations will allow an improvement in the model quality. In
\Cref{tab:share_conformer_blocks}, we begin by reviewing the model quality with a single physical conformer layer repeated different numbers of times. We observe that even with one physical conformer block (SL0-SL2), the WER decreases as the number of repetitions is increased (i.e., an increase in the number of virtual layers). With just $3$ repetitions of the single conformer layer (SL2), our model is able to decrease the WER by $-2.30$ and $-1.82$ for dev and test respectively. Even so, the WER rates are still large, so while we demonstrated that increasing the number of virtual layers can improve the model quality, the quality is still good enough. To improve this, we increase to $4$ physical conformer layers (SL3-SL6) which also brings our model size closer towards our goal of $5$M parameters. We find that repeating the conformer block transformations three times (SL5) reduces the WER by $-0.93$ and $-0.85$ for dev and test respectively compared to just one iteration of each conformer block layer. However, we find that further increasing to four repetitions per conformer block (SL6) begins to degrade our model quality. Thus, there is a limit to the number of times conformer blocks should be repeated.

\begin{table}[]
    \centering
    \begin{tabular}{|c|c|c|c|c|c|}
    \cline{2-6}
        \multicolumn{1}{c|}{} & \multirow{2}{*}{\thead{Non-Shared\\Modules}} & \multirow{2}{*}{\thead{Model\\Dim.}} & \multicolumn{2}{c|}{\thead{WER}} & \multirow{2}{*}{\thead{Model\\Size}} \\
    \cline{4-5}
        \multicolumn{1}{c|}{} & & & \thead{dev} & \thead{test} &  \\
\cline{2-6}\noalign{\vspace{2.5pt}}
    \hline
    SM0 & F.F. Start & 96 & 3.56 & 3.64 & 4.93M \\
    \hline
    SM1 & Attention & 128 & 3.19 & 3.48 & 4.99M \\
    \hline
    SM2 & Convolution & 136 & 3.13 & 3.35 & 5.03M \\
    \hline
    SM3 & F.F. End & 96 & 3.64 & 3.88 & 4.93M \\
    \hline
    \multirow{2}{*}{SM4} & \multirow{2}{*}{\makecell[c]{Attention\\+ Convolution}} & \multirow{2}{*}{120} & \multirow{2}{*}{3.22} & \multirow{2}{*}{3.36} & \multirow{2}{*}{5.03M} \\
    & & & & & \\
    \hline
    \end{tabular}
    \caption{Sharing conformer layers (i.e., $4$ Physical, $12$ Virtual), while unsharing specific modules.}
    \label{tab:shared_modules}
\vspace{-10pt}
\end{table}

\vspace{-3pt}
\subsection{Sharing Conformer Modules}
\vspace{-2pt}
Next, we dig into the structure of the conformer blocks to identify the major modules which we can either enable or disable sharing. 
In \Cref{tab:shared_modules}, we use the shared conformer block model with $4$ physical conformer layers repeated $3$ times (SL5) as our base model and then select certain conformer modules to disable sharing (i.e., unshare). By unsharing individual modules, the model size increases, and thus, we must reduce the internal model dimension to compensate. Unsharing the convolution layer (SM2) offers the lowest WER rates at $3.13$ and $3.35$ for dev and test respectively. However, it is interesting to observe that unsharing the feed forward start (SM0) and feed forward end (SM3) modules significantly increases the WER rates. We can attribute this to the fact that both feed forward modules are so large in size, and thus, by not sharing these modules, we must greatly reduce the size of the model weight dimensionality hyperparameter to compensate.

\begin{table}[]
    \centering
    \begin{tabular}{|c|c|c|c|c|c|}
\cline{2-6}
    \multicolumn{1}{c|}{} &\multicolumn{2}{c|}{\thead{Non-Shared Sub-Components}} & \multicolumn{2}{c|}{\thead{WER}} & \multirow{2}{*}{\thead{Model\\Size}} \\
\cline{2-5}
    \multicolumn{1}{c|}{} & \thead{Module} & \thead{Sub-Component} & \thead{dev} & \thead{test} &  \\
\cline{2-6}\noalign{\vspace{2.5pt}}
\hline
SC0 & F.F. Start & Linear (1) & 3.10 & 3.23 & 6.02M \\
\hline
SC1 & F.F. Start & Linear (2) & 3.16 & 3.20 & 6.02M \\
\hline
\hline
SC2 & Attention & Query & 3.24 & 3.37 & 5.01M \\
\hline
SC3 & Attention & Value & 3.23 & 3.40 & 5.01M \\
\hline
SC4 & Attention & Key & 3.09 & 3.29 & 5.01M \\
\hline
\hline
SC5 & Conv. & Pre-Conv. & 5.49 & 5.81 & 5.17M \\
\hline
SC6 & Conv. & Conv. & 3.02 & 3.16 & 5.35M \\
\hline
SC7 & Conv. & Post-Conv. & 3.37 & 3.71 & 5.01M \\
\hline
\hline
SC8 & F.F. End & Linear (1) & 2.99 & 3.18 & 6.02M \\
\hline
SC9 & F.F. End & Linear (2) & 3.08 & 3.30 & 6.02M \\
\hline
\hline
SC10 & All & Misc. Small & 2.95 & 3.28 & 5.36M \\
\hline
    \end{tabular}
    \caption{Effect of allowing certain conformer sub-components to be shared or not shared.}
    \label{tab:unshare_sub_weights}
\vspace{-3pt}
\end{table}

\vspace{-3pt}
\subsection{Sharing Sub-Components}
\vspace{-2pt}
To further our understanding of how sharing of different components affects quality, we next look at disabling sharing (i.e., unsharing) for specific sub-components within our model. Again, we leverage the best model from \Cref{tab:share_conformer_blocks} where we have $4$ physical layers repeated three times giving a total of $12$ virtual transformations (SL5) which achieved a WER of $3.20$ for a model of size $4.84$M. In \Cref{tab:unshare_sub_weights}, we unshare single weight variables at a time for each module. We see that unsharing these sub-components still keeps our model size close to $5$M parameters except in the case of the linear sub-components in both feed forward start and end modules (SC0, SC1, SC8, and SC9). In addition to unsharing the individual module sub-components which were shown in Fig.~\ref{fig:conformer_block_structure}, a number of other significantly smaller weights are also found within each module. These weights are small enough that they do not have a large impact on the overall model size (i.e., only an increase of $0.52$M parameters). Thus, we also evaluate unsharing these miscellaneous small weights as well (SC10). We can see that unsharing the convolution sub-components within the convolution module allows for the lowest WER (SC6) while unsharing the other sub-components in the conformer layer each result in an increase in the WER. For the attention module, unsharing both query (SC2) and value (SC3) results in similar WER to the original model, yet unsharing key (SC4) does see a decrease in WER, thus implying that the attention key sub-components contains important information for our model.

\begin{table}[]
    \centering
    \begin{tabular}{|c|c|c|c|c|c|c|}
\cline{2-7}
    \multicolumn{1}{c|}{} & \multicolumn{2}{c|}{\thead{\# Conformer Layers}} & \multirow{2}{*}{\thead{Rank (k)}} & \multicolumn{2}{c|}{\thead{WER}} & \multirow{2}{*}{\thead{Model\\Size}} \\
\cline{2-3}\cline{5-6}
    \multicolumn{1}{c|}{} & \thead{Physical} & \thead{Virtual} & & \thead{dev} & \thead{test} & \\
\cline{2-7}\noalign{\vspace{2.5pt}}
\hline
LR0 & 4 & 4 & N/A & 4.13 & 4.30 & 4.84M \\
\hline
LR1 & 8 & 8 & 50 & 3.46 & 3.69 & 5.04M \\
\hline
LR2 & 12 & 12 & 20 & 3.70 & 3.75 & 4.98M \\
\hline
LR3 & 16 & 16 & 6 & 3.81 & 4.05 & 5.00M \\
\hline
\hline
LRS0 & 8 & 16 & 50 & 3.14 & 3.36 & 5.04M \\
\hline
LRS1 & 8 & 24 & 50 & 2.99 & 3.23 & 5.04M \\
\hline
LRS2 & 8 & 32 & 50 & 2.88 & 3.25 & 5.04M \\
\hline
LRS3 & 8 & 40 & 50 & 2.84 & 2.98 & 5.04M \\
\hline
    \end{tabular}
    \caption{After applying low-rank architecture for feed forward modules. With and without sharing layers.}
    \label{tab:low_rank}
\vspace{-5pt}
\end{table}

\vspace{-3pt}
\subsection{Low-Rank (and Sharing)}
\vspace{-2pt}
Next we look towards low-rank architecture in \Cref{tab:low_rank}. By reducing the $k$, we can subsequently achieve an increase in the number of physical layers. As we can see, with $k=50$ (LR1), we are able to increase from $8$ physical layers compared to only $4$ when a low-rank architecture is not applied (LR0). We find that $k=50$ also decreases the WER of the model by $-0.67$ and $-0.61$ for dev and test respectively. However, while we expect increasing the number of physical conformer layers should improve the quality, we find that $k$ directly counteracts these WER improvements and thus while $k=20$ (LR2) and $k=6$ (LR3) achieve lower WER compared to the non low-rank architecture, they both perform worse than $k=50$ (LR1). Continuing with $k=50$ and the number of physical layers at $8$, we also apply our layer sharing technique\footnote{Preliminary results show only marginal improvements in combining low-rank and sub-component sharing due to the large search space.} to increase the number of virtual layers from $16$ (LRS0) up until $40$ (LRS3) by repeating each conformer layer. With this, we find WERs as low as $2.84$ and $2.98$ on dev and test are achievable at our bounds of $5$M parameters when repeating the $8$ physical conformer layers $5$ times each (LRS3).

\vspace{-3pt}
\subsection{Overview}
\vspace{-2pt}
Our overall best results for the evaluated methods are shown in \Cref{tab:overview}.
Each of our evaluated models was created based on an initial $14$M parameter model (B0) described in~\cite{conformer}.
We compare these models to a handcrafted  $5$M parameter model (B1) which was created by manually reducing hyperparameters (e.g., number of conformer blocks). 
The lowest overall WER was achieved by LRS3 because 
low-rank decomposition reduces the size of each physical conformer layer, thus allowing for a greater number of physical conformer layers while sharing layer weights through repeating offers an even greater number of virtual conformer layer transformations without increasing model size.
While reducing model size does increase WER compared to larger models, our goal in this work is to create a model which fits completely within TPU memory, thus offering low-power, always-on ASR. With this, we can handle most ASR tasks, while defer to a larger model only when necessary.

\begin{table}[]
    \centering
    \begin{tabular}{|l|c|c|c|}
    \hline
    \multicolumn{1}{|c|}{\multirow{2}{*}{\thead{Model}}} & \multicolumn{2}{c|}{\thead{WER}} & \multirow{2}{*}{\thead{Model\\Size}} \\
    \cline{2-3}
     & \thead{dev} & \thead{test} &  \\
    \hline
    \hline
    Conformer (S) \cite{conformer} (B0) & 2.18 & 2.53 & 14M \\
    \hline
    \hline
    Handcrafted (B1) & 3.53 (-0.00) & 3.72 (-0.00) & 4.9M \\
    \hline
    \hline
    Share Layers (SL5) & 3.20 (-0.33) & 3.45 (-0.27) & 4.84M \\
    \hline
    Share Modules (SM2) & 3.13 (-0.40) & 3.35 (-0.37) & 5.09M \\
    \hline
    Shared Sub-C. (SC10) & 2.95 (-0.58) & 3.28 (-0.44) & 5.36M \\
    \hline
    Low-Rank (LR2) & 3.46 (-0.07) & 3.69 (-0.03) & 5.04M \\
    \hline
    L.R. Share (LRS3) & 2.84 (-0.69) & 2.98 (-0.74) & 5.04M \\
    \hline
    \end{tabular}
    \caption{Overall best results for the evaluated compression methods.}
    \label{tab:overview}
\vspace{-10pt}
\end{table}

\vspace{-3pt}
\section{Conclusion}
\label{sec:conclusion}
\vspace{-2pt}

In this work, we propose to reduce the size of Conformer-based models through parameter weight reuse at four levels: (i) repeating conformer block layer transformations, (ii) sharing specific conformer modules, (iii) sharing or not sharing sub-components per conformer module, and (iv) sharing low-rank decomposed sub-weights. 
By sharing model weight across layers, we find that we can increase the number of virtual transformations of our input data without further increasing the size of our model and thus we can retain our model in-memory for always-on ambient ASR leveraging low-power and low-resource neural accelerators such as edge TPU hardware.
Through our evaluations, we find that sharing model weights and applying a low-rank Conformer architecture (LRS3) offers the greatest performance for our $5$M parameter models, achieving a WER of $2.84$ and $2.98$ for LibriSpeech dev-clean and test-clean respectively.

\vfill\pagebreak

\bibliographystyle{IEEEbib}
\bibliography{refs}

\begin{thebibliography}{10}
\providecommand{\url}[1]{#1}
\csname url@rmstyle\endcsname
\providecommand{\newblock}{\relax}
\providecommand{\bibinfo}[2]{#2}
\providecommand\BIBentrySTDinterwordspacing{\spaceskip=0pt\relax}
\providecommand\BIBentryALTinterwordstretchfactor{4}
\providecommand\BIBentryALTinterwordspacing{\spaceskip=\fontdimen2\font plus
\BIBentryALTinterwordstretchfactor\fontdimen3\font minus
  \fontdimen4\font\relax}
\providecommand\BIBforeignlanguage[2]{{%
\expandafter\ifx\csname l@#1\endcsname\relax
\typeout{** WARNING: IEEEtran.bst: No hyphenation pattern has been}%
\typeout{** loaded for the language `#1'. Using the pattern for}%
\typeout{** the default language instead.}%
\else
\language=\csname l@#1\endcsname
\fi
#2}}

\bibitem{alvarez2019end}
Raziel Alvarez and Hyun-Jin Park,
\newblock ``{End-to-end Streaming Keyword Spotting},''
\newblock in {\em ICASSP 2019-2019 IEEE International Conference on Acoustics,
  Speech and Signal Processing (ICASSP)}. IEEE, 2019, pp. 6336--6340.

\bibitem{ray2021listen}
Swayambhu~Nath Ray, Minhua Wu, Anirudh Raju, Pegah Ghahremani, Raghavendra
  Bilgi, Milind Rao, Harish Arsikere, Ariya Rastrow, Andreas Stolcke, and Jasha
  Droppo,
\newblock ``{Listen with Intent: Improving Speech Recognition with
  Audio-to-Intent Front-End},''
\newblock {\em arXiv preprint arXiv:2105.07071}, 2021.

\bibitem{antonini2019resource}
Mattia Antonini, Tran~Huy Vu, Chulhong Min, Alessandro Montanari, Akhil Mathur,
  and Fahim Kawsar,
\newblock ``{Resource Characterisation of Personal-Scale Sensing Models on Edge
  Accelerators},''
\newblock in {\em Proceedings of the First International Workshop on Challenges
  in Artificial Intelligence and Machine Learning for Internet of Things},
  2019, pp. 49--55.

\bibitem{conformer}
Anmol Gulati, James Qin, Chung{-}Cheng Chiu, Niki Parmar, Yu~Zhang, Jiahui Yu,
  Wei Han, Shibo Wang, Zhengdong Zhang, Yonghui Wu, and Ruoming Pang,
\newblock ``{Conformer: Convolution-augmented Transformer for Speech
  Recognition},''
\newblock in {\em Interspeech 2020, 21st Annual Conference of the International
  Speech Communication Association, Virtual Event, Shanghai, China, 25-29
  October 2020}, Helen Meng, Bo~Xu, and Thomas~Fang Zheng, Eds. 2020, pp.
  5036--5040, {ISCA}.

\bibitem{hubara2016binarized}
Itay Hubara, Matthieu Courbariaux, Daniel Soudry, Ran El-Yaniv, and Yoshua
  Bengio,
\newblock ``{Binarized Neural Networks},''
\newblock {\em Advances in neural information processing systems}, vol. 29,
  2016.

\bibitem{han2015learning}
Song Han, Jeff Pool, John Tran, and William Dally,
\newblock ``{Learning both Weights and Connections for Efficient Neural
  Network},''
\newblock {\em Advances in neural information processing systems}, vol. 28,
  2015.

\bibitem{hernandez2022wifi}
Steven~M Hernandez and Eyuphan Bulut,
\newblock ``{WiFi Sensing on the Edge: Signal Processing Techniques and
  Challenges for Real-World Systems},''
\newblock {\em IEEE Communications Surveys \& Tutorials}, 2022.

\bibitem{xu2022ultra}
Kunran Xu, Huawei Zhang, Yishi Li, Yuhao Zhang, Rui Lai, and Yi~Liu,
\newblock ``{An Ultra-low Power TinyML System for Real-time Visual Processing
  at Edge},''
\newblock {\em arXiv preprint arXiv:2207.04663}, 2022.

\bibitem{banbury2020benchmarking}
Colby~R Banbury, Vijay~Janapa Reddi, Max Lam, William Fu, Amin Fazel, Jeremy
  Holleman, Xinyuan Huang, Robert Hurtado, David Kanter, Anton Lokhmotov,
  et~al.,
\newblock ``{Benchmarking TinyML Systems: Challenges and Direction},''
\newblock {\em arXiv preprint arXiv:2003.04821}, 2020.

\bibitem{wu2021dynamic}
Zhaofeng Wu, Ding Zhao, Qiao Liang, Jiahui Yu, Anmol Gulati, and Ruoming Pang,
\newblock ``{Dynamic Sparsity Neural Networks for Automatic Speech
  Recognition},''
\newblock in {\em ICASSP 2021-2021 IEEE International Conference on Acoustics,
  Speech and Signal Processing (ICASSP)}. IEEE, 2021, pp. 6014--6018.

\bibitem{ding2021audio}
Shaojin Ding, Tianlong Chen, and Zhangyang Wang,
\newblock ``{Audio Lottery: Speech Recognition Made Ultra-Lightweight,
  Noise-Robust, and Transferable},''
\newblock in {\em International Conference on Learning Representations}, 2021.

\bibitem{zhou2018adaptive}
Yiren Zhou, Seyed-Mohsen Moosavi-Dezfooli, Ngai-Man Cheung, and Pascal
  Frossard,
\newblock ``{Adaptive Quantization for Deep Neural Network},''
\newblock in {\em Proceedings of the AAAI Conference on Artificial
  Intelligence}, 2018, vol.~32.

\bibitem{novac2021quantization}
Pierre-Emmanuel Novac, Ghouthi Boukli~Hacene, Alain Pegatoquet, Beno{\^\i}t
  Miramond, and Vincent Gripon,
\newblock ``{Quantization and Deployment of Deep Neural Networks on
  Microcontrollers},''
\newblock {\em Sensors}, vol. 21, no. 9, pp. 2984, 2021.

\bibitem{ding22c_interspeech}
Shaojin Ding, Phoenix Meadowlark, Yanzhang He, Lukasz Lew, Shivani Agrawal, and
  Oleg Rybakov,
\newblock ``{4-bit Conformer with Native Quantization Aware Training for Speech
  Recognition},''
\newblock in {\em Proc. Interspeech 2022}, 2022, pp. 1711--1715.

\bibitem{DBLP:conf/interspeech/CeruttiPBF19}
Gianmarco Cerutti, Rahul Prasad, Alessio Brutti, and Elisabetta Farella,
\newblock ``{Neural Network Distillation on IoT Platforms for Sound Event
  Detection},''
\newblock in {\em Interspeech 2019, 20th Annual Conference of the International
  Speech Communication Association, Graz, Austria, 15-19 September 2019},
  Gernot Kubin and Zdravko Kacic, Eds. 2019, pp. 3609--3613, {ISCA}.

\bibitem{yang2020model}
Ze~Yang, Linjun Shou, Ming Gong, Wutao Lin, and Daxin Jiang,
\newblock ``{Model Compression with Two-stage Multi-teacher Knowledge
  Distillation for Web Question Answering System},''
\newblock in {\em Proceedings of the 13th International Conference on Web
  Search and Data Mining}, 2020, pp. 690--698.

\bibitem{DBLP:journals/corr/abs-2207-00112}
Yen{-}Chang Hsu, Ting Hua, Sungen Chang, Qian Lou, Yilin Shen, and Hongxia Jin,
\newblock ``{Language model compression with weighted low-rank
  factorization},''
\newblock {\em CoRR}, vol. abs/2207.00112, 2022.

\bibitem{yu2017compressing}
Xiyu Yu, Tongliang Liu, Xinchao Wang, and Dacheng Tao,
\newblock ``{On Compressing Deep Models by Low Rank and Sparse
  Decomposition},''
\newblock in {\em Proceedings of the IEEE conference on computer vision and
  pattern recognition}, 2017, pp. 7370--7379.

\bibitem{wen2016learning}
Wei Wen, Chunpeng Wu, Yandan Wang, Yiran Chen, and Hai Li,
\newblock ``{Learning Structured Sparsity in Deep Neural Networks},''
\newblock {\em Advances in neural information processing systems}, vol. 29,
  2016.

\bibitem{schaefer2022edge}
Clemens~JS Schaefer, Siddharth Joshi, Shan Li, and Raul Blazquez,
\newblock ``{Edge Inference with Fully Differentiable Quantized Mixed Precision
  Neural Networks},''
\newblock {\em arXiv preprint arXiv:2206.07741}, 2022.

\bibitem{vaswani2017attention}
Ashish Vaswani, Noam Shazeer, Niki Parmar, Jakob Uszkoreit, Llion Jones,
  Aidan~N Gomez, {\L}ukasz Kaiser, and Illia Polosukhin,
\newblock ``{Attention is All you Need},''
\newblock {\em Advances in neural information processing systems}, vol. 30,
  2017.

\bibitem{dabre2019recurrent}
Raj Dabre and Atsushi Fujita,
\newblock ``{Recurrent Stacking of Layers for Compact Neural Machine
  Translation Models},''
\newblock in {\em Proceedings of the AAAI Conference on Artificial
  Intelligence}, 2019, vol.~33, pp. 6292--6299.

\bibitem{lan2019albert}
Zhenzhong Lan, Mingda Chen, Sebastian Goodman, Kevin Gimpel, Piyush Sharma, and
  Radu Soricut,
\newblock ``{ALBERT: A Lite BERT for Self-supervised Learning of Language
  Representations},''
\newblock {\em arXiv preprint arXiv:1909.11942}, 2019.

\bibitem{xue2013restructuring}
Jian Xue, Jinyu Li, and Yifan Gong,
\newblock ``{Restructuring of Deep Neural Network Acoustic Models with Singular
  Value Decomposition},''
\newblock in {\em Interspeech}, 2013, pp. 2365--2369.

\bibitem{panayotov2015librispeech}
Vassil Panayotov, Guoguo Chen, Daniel Povey, and Sanjeev Khudanpur,
\newblock ``{Librispeech: An ASR corpus based on public domain audio books},''
\newblock in {\em 2015 IEEE international conference on acoustics, speech and
  signal processing (ICASSP)}, 2015.

\bibitem{han2020contextnet}
Wei Han, Zhengdong Zhang, Yu~Zhang, Jiahui Yu, Chung-Cheng Chiu, James Qin,
  Anmol Gulati, Ruoming Pang, and Yonghui Wu,
\newblock ``{ContextNet: Improving Convolutional Neural Networks for Automatic
  Speech Recognition with Global Context},''
\newblock {\em Proc. Interspeech 2020}, pp. 3610--3614, 2020.

\end{thebibliography}

\end{document}